\documentclass[twocolumn,aps,pra,showpacs]{revtex4}
\usepackage[T1]{fontenc}
\usepackage[latin9]{inputenc}
\usepackage{amsmath}
\usepackage{amssymb}
\usepackage{graphicx}
\usepackage{esint}

\makeatletter

\providecommand{\tabularnewline}{\\}

\@ifundefined{textcolor}{}
{%
 \definecolor{BLACK}{gray}{0}
 \definecolor{WHITE}{gray}{1}
 \definecolor{RED}{rgb}{1,0,0}
 \definecolor{GREEN}{rgb}{0,1,0}
 \definecolor{BLUE}{rgb}{0,0,1}
 \definecolor{CYAN}{cmyk}{1,0,0,0}
 \definecolor{MAGENTA}{cmyk}{0,1,0,0}
 \definecolor{YELLOW}{cmyk}{0,0,1,0}
}

\@ifundefined{textcolor}{}{%
 \definecolor{BLACK}{gray}{0}
 \definecolor{WHITE}{gray}{1}
 \definecolor{RED}{rgb}{1,0,0}
 \definecolor{GREEN}{rgb}{0,1,0}
 \definecolor{BLUE}{rgb}{0,0,1}
 \definecolor{CYAN}{cmyk}{1,0,0,0}
 \definecolor{MAGENTA}{cmyk}{0,1,0,0}
 \definecolor{YELLOW}{cmyk}{0,0,1,0}
}

\makeatother

\begin{document}

\title{Fermi polaron in a one-dimensional quasi-periodic optical lattice:
the simplest many-body localization challenge }

\author{Hui Hu$^{1,2}$}

\author{An-Bang Wang$^{1}$}

\author{Su Yi$^{1}$}

\author{Xia-Ji Liu$^{2}$}

\affiliation{$^{1}$Institute of Theoretical Physics, Chinese Academy of Sciences,
Beijing 100190, P. R. China}

\affiliation{$^{2}$Centre for Quantum and Optical Science, Swinburne University
of Technology, Melbourne 3122, Australia}

\date{\today}
\begin{abstract}
We theoretically investigate the behavior of a moving impurity immersed
in a sea of fermionic atoms that are confined in a quasi-periodic
(bichromatic) optical lattice, within a standard variational approach.
We consider both repulsive and attractive contact interactions for
such a simplest many-body localization problem of Fermi polarons.
The variational approach enables us to access relatively large systems
and therefore may be used to understand many-body localization in
the thermodynamic limit. The energy and wave-function of the polaron
states are found to be strongly affected by the quasi-random lattice
potential and their experimental measurements (i.e., via radio-frequency
spectroscopy or quantum gas microscope) therefore provide a sensitive
way to underpin the localization transition. We determine a phase
diagram by calculating two critical quasi-random disorder strengths,
which correspond to the onset of the localization of the ground-state
polaron state and the many-body localization of all polaron states,
respectively. Our predicted phase diagram could be straightforwardly
examined in current cold-atom experiments. 
\end{abstract}


\maketitle

\section{Introduction}

Anderson localization of interacting disordered systems - a phenomenon
referred to as many-body localization (MBL) - has received intense
attention over the past few years \cite{Nandkishore2015,Altman2015}.
Earlier studies focus on condensed matter systems, where a uniformly
distributed white-noise disorder potential is often adopted to carry
out perturbative analyses in the presence of weak interactions \cite{Basko2006,Aleiner2010}
or numerical simulations with strong interactions \cite{Oganesyan2007,Pal2010,Bardarson2012,Mondaini2015}.
Recent experimental advances in ultracold atoms provide a new paradigm
to explore MBL \cite{Schreiber2015,Bordia2015}. In these experiments,
a quasi-periodic bichromatic optical lattice has been used, leading
to a quasi-random disorder potential \cite{Harper1955,Aubry1980}.
The interatomic interaction and dimensionality of the system can be
tuned at will, with unprecedented accuracy \cite{Bloch2008}.

It is well-known that Anderson localization occurs not only in the
ground state of the system but also in highly excited states \cite{Evers2008}.
In the presence of interactions, this fundamental feature makes both
theoretical and experimental investigations of MBL extremely challenging
\cite{Nandkishore2015,Altman2015}. To understand the localization
of highly excited states, most theoretical studies of interacting
disorder systems rely on the full exact diagonalization of the model
Hamiltonian and therefore the size of the system is severely restricted
\cite{Oganesyan2007,Pal2010,Mondaini2015}. On the other hand, in
the recent two cold-atom experiments, only the localization of a particular
type of (excited) states, i.e., a charge density wave state, has been
examined \cite{Schreiber2015,Bordia2015}.

\begin{figure}
\begin{centering}
\includegraphics[clip,width=0.45\textwidth]{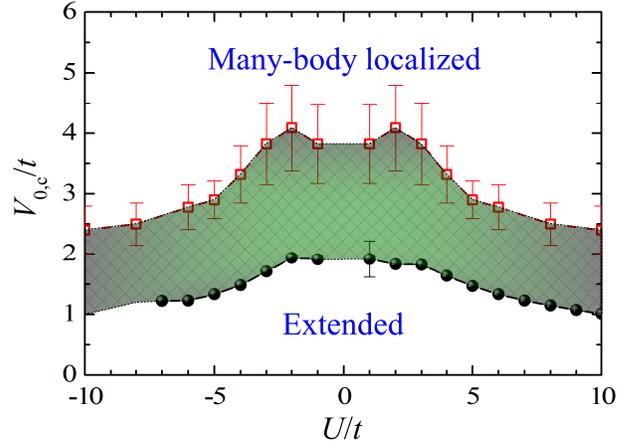} 
\par\end{centering}

\caption{(Color online) Phase diagram of a fermionic polaron in quasi-random
lattice potentials. The full circles with solid line show the threshold
of entering a localized state for the ground-state polaron. The empty
squares with dashed line give the critical disorder strength, above
which \emph{all} polaron states become localized. The error bars indicate
the estimated uncertainty. The system has an unpure many-body energy
spectrum in the shaded area, i.e., extended polaron states can coexist
with localized polaron states. The average filling factor of fermionic
atoms is $\left\langle \hat{n}\right\rangle =1/2$.}

\label{fig1} 
\end{figure}

In this work, motivated by those rapid experimental advances, we propose
to experimentally explore (arguably) the simplest case of MBL: a moving
impurity immersed in a sea of non-interacting fermionic atoms. The
latter is subjected to a one-dimensional (1D) quasi-periodic optical
lattice \cite{Harper1955} and experiences Anderson localization when
the disorder strength is sufficiently strong \cite{Aubry1980}. There
is a contact interaction between impurity and fermionic atoms. The
motion of impurity - or more precisely a Fermi polaron \cite{Chevy2006,Lobo2006,Achirotzek2009,Wenz2013,Massignan2014}
- is therefore affected by the localization properties of fermionic
atoms. The proposed system has several advantages to address the MBL
phenomenon. Experimentally, it seems easier to measure a Fermi polaron.
Its energy might be determined by using radio-frequency spectroscopy
\cite{Achirotzek2009} while its wave-function might be identified
from the \textit{in situ} density profile through the recently developed
quantum gas microscope for fermionic atoms \cite{Cheuk2015,Parsons2015,Omran2015}.
Theoretically, we have well-controlled approximations to handle the
Fermi polaron problem \cite{Chevy2006}, even in the limit of very
strong interactions \cite{Lobo2006,Doggen2014}, which make it feasible
to access large systems as those experimentally explored. Therefore,
we may underpin a phase diagram of MBL in the thermodynamic limit.

Our main result is briefly summarized in Fig. \ref{fig1}. We determine
two critical quasi-random disorder strengths within a variational
approach, by taking into account the dominant single particle-hole
excitation above the Fermi sea \cite{Chevy2006}. The first critical
field (circles with solid line) corresponds to the onset of the Anderson
localization of the ground-state polaron state. While at the second
critical field (empty squares with dashed line), all polaron states
become localized. This gives rise to a complete MBL phase diagram
of Fermi polarons. It is amazing that even the interaction experienced
by a \emph{single} impurity can dramatically lead to the appearance
of MBL. Our predicted phase diagram could be easily examined in current
cold-atom experiments \cite{Schreiber2015,Bordia2015}.

\section{Model Hamiltonian and variational approach}

A moving Fermi polaron in a 1D quasi-disordered lattice of length
$L$ can be described by the model Hamiltonian \cite{Schreiber2015,Bordia2015},
\begin{eqnarray}
{\cal H} & = & \sum_{n=0}^{L-1}\left[\left(-t_{c}\hat{c}_{n}^{\dagger}\hat{c}_{n+1}+\text{H.c.}\right)+\left(-t_{d}\hat{d}_{n}^{\dagger}\hat{d}_{n+1}+\text{H.c.}\right)\right.\nonumber \\
 &  & +\left.U\hat{c}_{n}^{\dagger}\hat{c}_{n}\hat{d}_{n}^{\dagger}\hat{d}_{n}+V_{0}\cos\left(2\pi n\beta+\theta\right)\hat{c}_{n}^{\dagger}\hat{c}_{n}\right],\label{eq:PolaronHami}
\end{eqnarray}
where $\hat{c}_{n}$ and $\hat{d}_{n}$ are the annihilation field
operators for fermionic atoms and impurity, respectively. For atoms,
we consider the half-filling case that corresponds to a chemical potential
$\mu=0$. In contrast, there is only one impurity which creates a
single Fermi polaron. $t_{c}$ and $t_{d}$ are the hopping amplitudes
and without loss of generality, we take equal mass for atoms and impurity
and hence $t_{c}=t_{d}=t$. We use a periodic boundary condition,
which means $\hat{c}_{L}=\hat{c}_{0}$ and $\hat{d}_{L}=\hat{d}_{0}$.
For single-component fermionic atoms, the interatomic interaction
is of $p$-wave characteristic and is generally very weak. Thus, we
assume an $s$-wave interaction between fermionic atoms and the impurity
only, with the interaction strength $U$ being either repulsive ($U>0$)
or attractive ($U<0$). The last term with the potential $V_{0}\cos(2\pi\beta n+\theta)$
in the Hamiltonian describes the quasi-periodic superlattice experienced
by fermionic atoms \cite{Schreiber2015,Bordia2015}. We assume that
the impurity does not feel the quasi-periodic potential and, without
the impurity-atom interaction, can move freely through the lattice.
In the presence of the interaction, the motion of impurity or Fermi
polaron therefore provides a sensitive probe of the underlying localization
properties of the Fermi sea background.

In the quasi-disorder potential, the irrational number $\beta$ and
phase offset $\theta$ are determined by the experimental bichromatic
lattice setup \cite{Schreiber2015,Bordia2015}. However, from a theoretical
point of view, their detailed values are irrelevant \cite{Kohmoto1983}.
Hereafter, for definiteness we take $\beta=(\sqrt{5}-1)/2$, the inverse
of the golden mean, and $\theta=0$, unless specifically noted. To
increase the numerical stability, we further approximate $\beta$
as the limit of a continued fraction, $\beta\simeq F_{l-1}/F_{l}$
\cite{Dufour2012}, where $F_{l}$ are Fibonacci numbers (i.e., $F_{0}=F_{1}=1$
and $F_{l+1}=F_{l}+F_{l-1}$) and $l$ is a sufficiently large integer.
We minimize the finite-size effect by taking the length of the lattice
$L=F_{l}$ \cite{Dufour2012}.

In the absence of the impurity or the interaction, the model Hamiltonian
reduces to the well-known Aubry-André-Harper (AAH) model \cite{Harper1955,Aubry1980}.
Fermionic atoms experience Anderson localization at the critical point
$V_{c}^{(0)}=2t$, at which all the single-particle states of atoms
are multifractal \cite{Kohmoto1983}. If $V_{0}<V_{c}^{(0)}$, all
the states are extended. Otherwise ($V_{0}>V_{c}^{(0)}$), all the
states are exponentially localized \cite{Aubry1980}. Here, we address
the problem of how the behavior of the Fermi polaron is affected by
the localization properties of fermionic atoms, due to the impurity-atom
interaction.

\subsection{Variational approach with one particle-hole excitation}

To solve the Fermi polaron problem, we use the standard variational
approach within the approximation of considering only a \emph{single}
particle-hole excitation, as proposed by Chevy \cite{Chevy2006}.
This approach is known to provide an accurate zero-temperature description
of the equation of state and of the dynamics of the system in a reasonably
long time-scale \cite{Massignan2014,Doggen2014}. Let us consider
a Fermi sea of fermionic atoms, occupied up to the chemical potential
$\mu=0$ (i.e., at half-filling with $\left\langle \hat{n}\right\rangle =\sum_{n}\left\langle \hat{c}{}_{n}^{\dagger}\hat{c}_{n}\right\rangle =1/2$):
\begin{equation}
\left|\text{FS}\right\rangle =\prod\limits _{E_{\eta}<0}\hat{c}_{\eta}^{\dagger}\left|\text{vac}\right\rangle ,
\end{equation}
where $E_{\eta}$ is the energy level of the AAH model for fermionic
atoms in the quasi-random lattice and $\hat{c}_{\eta}$ is the corresponding
field operator. The level index $\eta$ of single-particle states
runs from $0$ to $\eta_{F}-1$, where $\eta_{F}$ is the first energy
level that satisfies $E_{\eta_{F}}>0$, and finally to $L-1$. For
a large lattice size $L\gg1$, we would have $\eta_{F}\simeq L/2$.
For the numerical convenience, we shall always take 
\begin{equation}
\eta_{F}=\frac{L}{2}.
\end{equation}
By slightly generalizing Chevy's variational ansatz \cite{Chevy2006},
a Fermi polaron in disordered potentials can be described by the following
approximate many-body wave-function, 
\begin{equation}
\left|\text{P}\right\rangle =\sum\limits _{n}z_{n}\hat{d}_{n}^{\dagger}\left|\text{FS}\right\rangle +\sum\limits _{n,\eta_{h},\eta_{p}}\alpha_{n}\left(\eta_{h},\eta_{p}\right)\hat{d}_{n}^{\dagger}\hat{c}_{\eta_{p}}^{\dagger}\hat{c}_{\eta_{h}}\left|\text{FS}\right\rangle ,
\end{equation}
where $z_{n}$ corresponds to the residue of the polaron at each lattice
site $n\in[0,L-1]$. The second term with amplitude $\alpha_{n}\left(\eta_{h},\eta_{p}\right)$
describes the single particle-hole excitation, for which the level
index $\eta_{h}$ (for hole excitation) and $\eta_{p}$ (for particle
excitation) satisfy 
\begin{equation}
0\leq\eta_{h}\leq\eta_{F}-1<\eta_{p}\leq L-1.
\end{equation}
We note that, in the absence of the quasi-random lattice, momentum
is a good quantum number and the level index $\eta$ will then simply
be replaced by momentum. In that case, we can use momentum conservation
to greatly simplify the wave-function so that the amplitude $\alpha_{n}(\eta_{h},\eta_{p})$
depends only on a momentum difference and Chevy's variational ansatz
is then recovered \cite{Chevy2006}. The loss of periodicity means
that we may have to restrict the length of the system $L$ to a reasonably
large value.

\subsection{The dimension of the polaron Hilbert space}

Let us now count how many states are there in the polaron variational
wave-function. The site index $n$ takes $L$ values, $\eta_{h}$
runs from $0$ to $\eta_{F}-1$, and finally $\eta_{p}$ takes $L-\eta_{F}$
values. This means that we should have a Hilbert space with dimension,
\begin{equation}
D=L\left[1+\eta_{F}\left(L-\eta_{F}\right)\right]\simeq\frac{L^{3}}{4},
\end{equation}
where we have used $\eta_{F}=L/2$. Thus, we obtain, with increasing
$l$, $D_{l=6}\simeq549$, $D_{l=7}\simeq2,315$, $D_{l=8}\simeq9,826$,
$D_{l=9}\simeq41,593$, $D_{l=10}\simeq176,242$, and $D_{l=11}\simeq746,496$,
as listed in Table \ref{table1}.

\begin{table}
\begin{centering}
\begin{tabular}{|c|c|c|c|c|c|c|}
\hline 
$l$  & 6  & 7  & 8  & 9  & 10  & 11\tabularnewline
\hline 
\hline 
$L=F_{l}$  & 13  & 21  & 34  & 55  & 89  & 144\tabularnewline
\hline 
$D$  & 549  & 2,315  & 9,826  & 41,593  & 176,242  & 746,496\tabularnewline
\hline 
\end{tabular}
\par\end{centering}

\protect\protect\caption{The length of the system $L=F_{l}$ and the dimension of the polaron
Hilbert space $D$ considered in our numerical calculations.}

\label{table1} 
\end{table}

\subsection{Diagonalization solution of polaron states}

A direct and convenient way to solve the variational parameters $z_{n}$
and $\alpha_{n}\left(\eta_{h},\eta_{p}\right)$ is to diagonalize
the model Hamiltonian Eq. (\ref{eq:PolaronHami}) in the Hilbert space
expanded by the states, $\left|i\right\rangle =d_{n}^{\dagger}\left|\text{FS}\right\rangle $
or $\left|i\right\rangle =d_{n}^{\dagger}c_{\eta_{p}}^{\dagger}c_{\eta_{h}}\left|\text{FS}\right\rangle $,
where the index $i$ (or $j$ used later) runs from $1$ to $D$.
Thus, we obtain the following three kinds of matrix elements $\mathcal{H}_{ij}$:\begin{widetext}
\begin{equation}
\left\langle \text{FS}\right|d_{n}{\cal H}d_{n^{\prime}}^{\dagger}\left|\text{FS}\right\rangle =\delta_{nn^{\prime}}\left[E_{FS}+U\sum\limits _{E_{\eta}<0}\left|u_{\eta n}\right|^{2}\right]-t\delta_{n\pm1,n^{\prime}},
\end{equation}
\begin{equation}
\left\langle \text{FS}\right|d_{n}{\cal H}d_{n^{\prime}}^{\dagger}c_{\eta_{p}^{\prime}}^{\dagger}c_{\eta_{h}^{\prime}}\left|\text{FS}\right\rangle =U\delta_{nn^{\prime}}u_{\eta_{p}^{\prime}n}\left(u_{\eta_{h}^{\prime}n}\right)^{*},
\end{equation}
and 
\begin{eqnarray}
\left\langle \text{FS}\right|c_{\eta_{h}}^{\dagger}c_{\eta_{p}}d_{n}{\cal H}d_{n^{\prime}}^{\dagger}c_{\eta_{p}^{\prime}}^{\dagger}c_{\eta_{h}^{\prime}}\left|\text{FS}\right\rangle  & = & \left[\delta_{nn^{\prime}}\left(E_{\eta_{p}}-E_{\eta_{h}}+E_{FS}+U\sum\limits _{E_{\eta}<0}\left|u_{\eta n}\right|^{2}\right)-t\delta_{n\pm1,n^{\prime}}\right]\delta_{\eta_{p}\eta_{p}^{\prime}}\delta_{\eta_{h}\eta_{h}^{\prime}}\nonumber \\
 &  & +U\delta_{nn^{\prime}}\left[\delta_{\eta_{h}\eta_{h}^{\prime}}\left(u_{\eta_{p}n}\right)^{*}u_{\eta_{p}^{\prime}n}-\delta_{\eta_{p}\eta_{p}^{\prime}}u_{\eta_{h}n}\left(u_{\eta_{h}^{\prime}n}\right)^{*}\right].
\end{eqnarray}
\end{widetext}In the above expressions, 
\begin{equation}
E_{FS}\equiv\sum\limits _{E_{\eta}<0}E_{\eta}
\end{equation}
is the energy of the Fermi sea of fermionic atoms and $u_{\eta n}$
is the wave-function of the single-particle state $\eta$ of atoms,
obtained by solving the AAH Hamiltonian, i.e., 
\begin{equation}
\hat{c}_{n}=\sum_{\eta}u_{\eta n}\hat{c}_{\eta}.
\end{equation}
By appropriately arranging the order of the variational states, the
matrix element $\mathcal{H}_{ij}$ can be easily calculated. The resulting
large and sparse matrix can be partially or fully diagonalized by
using standard numerical subroutines, leading directly to $z_{n}$
and $\alpha_{n}\left(\eta_{h},\eta_{p}\right)$ of the ground state
and excited states of the polaron.

\subsection{Variational minimization of the ground-state polaron state}

Alternatively, for the ground-state of the polaron, we may determine
$z_{n}$ and $\alpha_{n}\left(\eta_{h},\eta_{p}\right)$ by minimizing
$\left\langle \text{P}\right|{\cal H}\left|\text{P}\right\rangle $
\cite{Chevy2006}, under the normalization condition, 
\begin{equation}
\sum_{n}z_{n}^{2}+\sum\limits _{n,\eta_{h},\eta_{p}}\alpha_{n}^{2}\left(\eta_{h},\eta_{p}\right)=1.
\end{equation}
By taking some straightforward calculations, it is easy to obtain
that\begin{widetext}, 
\begin{eqnarray}
\left\langle \text{P}\right|{\cal H}\left|\text{P}\right\rangle  & = & -2t\sum_{n}z_{n}z_{n+1}+\sum_{n}z_{n}^{2}\left(E_{FS}+U\sum\limits _{E_{\eta}<0}\left|u_{\eta n}\right|^{2}\right)-2t\sum\limits _{n,\eta_{h},\eta_{p}}\alpha_{n}\left(\eta_{h},\eta_{p}\right)\alpha_{n+1}\left(\eta_{h},\eta_{p}\right)\nonumber \\
 &  & +\sum\limits _{n,\eta_{h},\eta_{p}}z_{n}\alpha_{n}\left(\eta_{h},\eta_{p}\right)\left(E_{\eta_{p}}-E_{\eta_{h}}+E_{FS}+U\sum\limits _{E_{\eta}<0}\left|u_{\eta n}\right|^{2}\right)+2U\sum\limits _{n,\eta_{h},\eta_{p}}z_{n}\alpha_{n}\left(\eta_{h},\eta_{p}\right)u_{\eta_{h}n}u_{\eta_{p}n}\nonumber \\
 &  & +U\sum\limits _{n,\eta_{h},\eta_{p},\eta_{p}^{\prime}}\alpha_{n}\left(\eta_{h},\eta_{p}\right)\alpha_{n}\left(\eta_{h},\eta_{p}^{\prime}\right)u_{\eta_{p}n}u_{\eta_{p}^{\prime}n}-U\sum\limits _{n,\eta_{h},\eta_{p},\eta_{h}^{\prime}}\alpha_{n}\left(\eta_{h},\eta_{p}\right)\alpha_{n}\left(\eta_{h}^{\prime},\eta_{p}\right)u_{\eta_{h}n}u_{\eta_{h}^{\prime}n},
\end{eqnarray}
where we have used the fact that the coefficients $u_{\eta n}$ are
real. The minimization of $\left\langle \text{P}\right|{\cal H}\left|\text{P}\right\rangle $
then leads to the following two coupled equations \cite{Chevy2006}:
\begin{equation}
0=-t\left(z_{n-1}+z_{n+1}\right)+\left(E_{FS}+U\sum\limits _{E_{\eta}<0}\left|u_{\eta n}\right|^{2}-\lambda\right)z_{n}+U\sum\limits _{\eta_{h},\eta_{p}}\alpha_{n}\left(\eta_{h},\eta_{p}\right)u_{\eta_{h}n}u_{\eta_{p}n}
\end{equation}
and 
\begin{eqnarray}
0 & = & z_{n}Uu_{\eta_{h}n}u_{\eta_{p}n}+\alpha_{n}\left(\eta_{h},\eta_{p}\right)\left(E_{\eta_{p}}-E_{\eta_{h}}+E_{FS}+U\sum\limits _{E_{\eta}<0}\left|u_{\eta n}\right|^{2}-\lambda\right)\nonumber \\
 &  & -t\left[\alpha_{n-1}\left(\eta_{h},\eta_{p}\right)+\alpha_{n+1}\left(\eta_{h},\eta_{p}\right)\right]+U\sum\limits _{\eta_{p}^{\prime}}\alpha_{n}\left(\eta_{h},\eta_{p}^{\prime}\right)u_{\eta_{p}n}u_{\eta_{p}^{\prime}n}-U\sum\limits _{\eta_{h}^{\prime}}\alpha_{n}\left(\eta_{h}^{\prime},\eta_{p}\right)u_{\eta_{h}n}u_{\eta_{h}^{\prime}n}.
\end{eqnarray}
\end{widetext}Here, $\lambda$ is a multiplier used to remove the
normalization constraint for the variational parameters. In the limit
of weak interactions, $\left|U\right|\ll t$, we may use the above
coupled equations to have a perturbative solution for $z_{n}$ and
$\alpha_{n}\left(\eta_{h},\eta_{p}\right)$.

\subsection{Properties of a polaron state}

The total residue of a polaron state is given by \cite{Chevy2006},
\begin{equation}
Z=\sum_{n}z_{n}^{2}.
\end{equation}
It seems reasonable to define a (normalized) wave-function for a Fermi
polaron, 
\begin{equation}
\psi_{n}=\frac{z_{n}}{\sqrt{Z}}.
\end{equation}
Moreover, the energy of the polaron $E$ can be written in relative
to its non-interacting counterpart, 
\begin{equation}
E_{P}=E-E_{FS}-E_{\textrm{imp}}^{(0)},
\end{equation}
where $E_{\textrm{imp}}^{(0)}=-2t_{d}=-2t$ is the lowest energy level
of the impurity with the Hamiltonian ${\cal H}_{d}^{(0)}=\sum_{n}(-t_{d}d_{n}^{\dagger}d_{n+1}+\textrm{H.c.})$.
By considering the scaling behavior of $E_{P}$ and its effective
wave-function $\psi_{n}$, as a function of the rational index $l$,
we may determine the localization property of the Fermi polaron \cite{Kohmoto1983}.

To characterize the localization transition of the \emph{ground-state}
polaron, it is convenient to use the inverse participation ratio (IPR)
defined by, 
\begin{equation}
\alpha_{\text{IPR}}=\sum_{n=0}^{L-1}\left|\psi_{n}\right|^{4}.
\end{equation}
For an extended state, we anticipate $\alpha_{\text{IPR}}$ $\sim1/L$,
while for a localized state, $\alpha_{\text{IPR}}$ converges to a
finite value at the order of $O(1)$. Near the localization transition
point, with increasing disorder strength, a sharp increase would appear
in $\alpha_{\text{IPR}}$.

The IPR is not a sensitive indicator for determining the localization
properties of \emph{excited} polaron states or MBL. In this case,
the system size $L$ is not allowed to take large values since the
information of all excited states is needed. As a result, one can
hardly carry out the scaling analysis of all excited states by increasing
the rational index $l$. It turns out to be more useful to consider
the statistics of the many-body energy spectrum, as suggested by Oganesyan
and Huse \cite{Oganesyan2007}. That is, the energy level spacing
of the many-body system has different probability distribution across
the MBL transition. Numerically, we may calculate the dimensionless
ratio between the smallest and largest adjacent energy gaps \cite{Oganesyan2007,Mondaini2015},
\begin{equation}
0\leqslant r_{n}=\frac{\min\left\{ \delta_{n},\delta_{n-1}\right\} }{\max\left\{ \delta_{n},\delta_{n-1}\right\} }\leqslant1,
\end{equation}
where $\delta_{n}=E_{n}-E_{n-1}\geqslant0$ and $\left\{ E_{n}\right\} $
is the ascending ordered list of the many-body energy levels. In the
extended phase, the ratio satisfies a Wigner-Dyson distribution (for
the Gaussian orthogonal ensemble, GOE) and the averaged ratio is,
\begin{equation}
\left\langle r_{n}\right\rangle _{\textrm{WD}}\simeq0.536.
\end{equation}
While in the MBL phase, the ratio follows a Poisson distribution $P_{\textrm{P}}(r)=2/(1+r)^{2}$,
with an averaged ratio, 
\begin{equation}
\left\langle r_{n}\right\rangle _{\textrm{P}}=2\ln2-1\simeq0.386.
\end{equation}

\section{Results and discussions}

\subsection{The ground-state polaron}

\begin{figure}
\begin{centering}
\includegraphics[clip,width=0.48\textwidth]{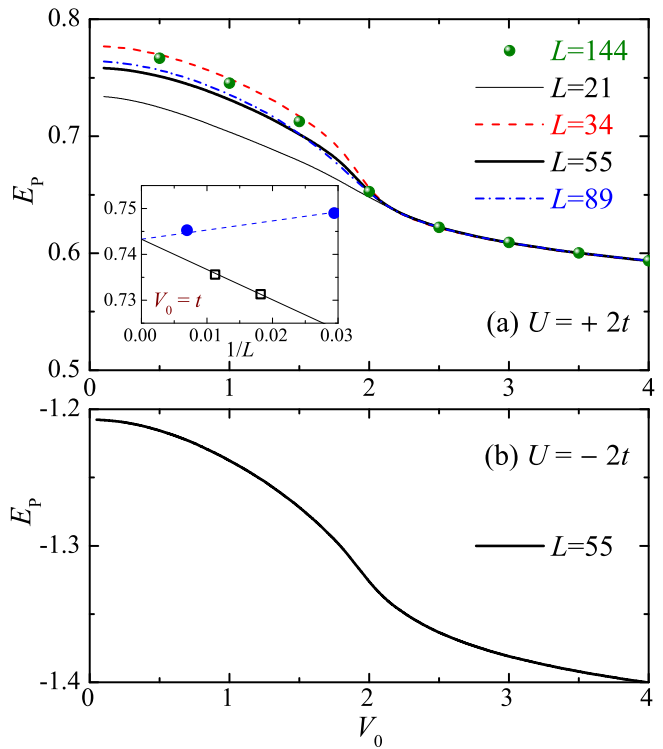} 
\par\end{centering}

\caption{(Color online) The ground-state energy of repulsive (a) and attractive
polarons (b) as a function of the disorder strength, at the interaction
strength $\left|U\right|=2t$. In (a), we check the dependence of
the polaron energy on the length of the system. The inset shows the
$1/L$ dependence of the polaron energy at a weak disorder strength
$V_{0}=t$. By extrapolating to $1/L=0$, we obtain $E_{P}(V_{0}=t)\simeq0.744t$
in the thermodynamic limit. In contrast, in the localized phase, the
length dependence is extremely weak. The average filling factor of
fermionic atoms is $\left\langle \hat{n}\right\rangle =1/2$.}

\label{fig2} 
\end{figure}

Figure \ref{fig2} reports the energy of the ground-state polaron
at an intermediate onsite interaction strength $\left|U\right|=2t$
for the system length up to $L=144$. Both the energy of repulsive
and attractive polarons decreases with increasing quasi-random disorder
strength. However, the length dependence of the polaron energy is
very different for weak and strong disorder. In the former case, the
finite-size effect is pronounced. Numerically we find that the finite-size
correction to energy is approximately proportional to $1/L$, as shown
in the inset at the disorder strength $V_{0}=t$, with a coefficient
that depends on the parity of the length. Thus, the polaron energy
approach its thermodynamic limit from above or below for even or odd
system length, respectively. In contrast, at strong disorder (i.e.,
$V_{0}>2t$), the polaron energy essentially does not depend on the
length.

\begin{figure}
\begin{centering}
\includegraphics[clip,width=0.48\textwidth]{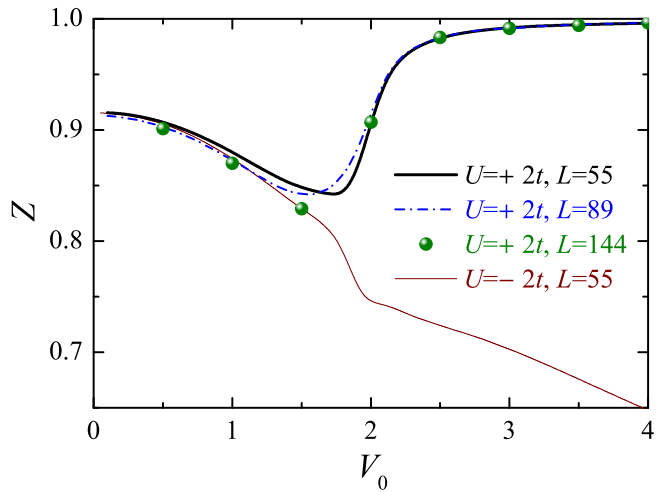} 
\par\end{centering}

\protect\protect\caption{(Color online) The residue of repulsive (thick solid line, dashed
line and circles) and attractive polarons (thin solid line) as a function
of the disorder strength, at the interaction strength $\left|U\right|=2t$.
We take an average filling factor of fermionic atoms $\left\langle \hat{n}\right\rangle =1/2$.}

\label{fig3} 
\end{figure}

The residue of the ground-state polaron similarly shows different
length dependence at weak and strong disorder, as illustrated in Fig.
\ref{fig3}. Furthermore, it is interesting that the behavior of the
residue is also affected by the sign of the impurity-atom interaction.
While the residues of both repulsive and attractive polarons initially
decrease with increasing disorder strength, beyond a threshold $V_{0,c}\thicksim2t$,
the residue of the repulsive polaron saturates to unity and that of
the attractive polaron continues to decrease. Therefore, for a repulsive
polaron, the impurity will finally be isolated by strong disorder.
In contrast, for an attractive polaron, the impurity will bind more
tightly with surrounding fermionic atoms in the strong disorder limit.
In other words, the formation of a molecule is favored at strong disorder.

\subsection{Localization of the ground-state polaron}

The different finite-size dependence of the polaron energy and residue
at weak and strong disorder indicates that there is a localization
transition of the ground-state polaron, which we now characterize
quantitatively by using the IPR.

\begin{figure}
\begin{centering}
\includegraphics[clip,width=0.48\textwidth]{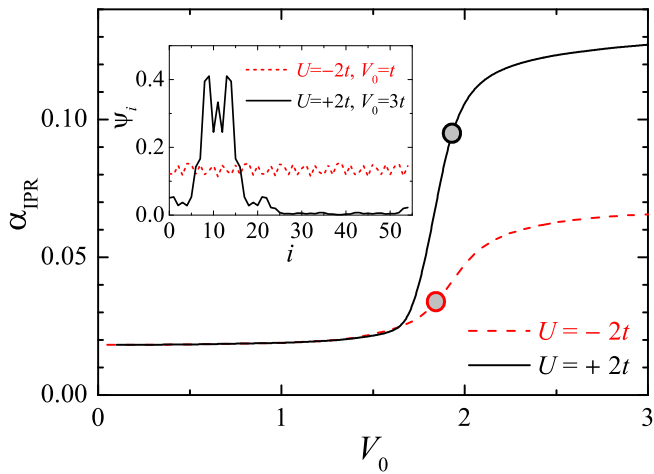} 
\par\end{centering}

\protect\protect\caption{(Color online) The inverse participation ratio of repulsive (solid
line) and attractive polarons (dashed line) as a function of the disorder
strength, at the interaction strength $\left|U\right|=2t$. The circles
indicate the inflection point of the curve (i.e., the threshold for
the localization of the ground-state polaron). The inset shows the
wave-function of the ground-state polaron $\psi_{i}$ at $\theta=2\pi/5$.
The other parameters are the same as in Fig. \ref{fig3}.}

\label{fig4} 
\end{figure}

Figure \ref{fig4} presents the disorder dependence of $\alpha_{\text{IPR}}$
of repulsive and attractive polarons at the interaction strength $\left|U\right|=2t$.
As anticipated, there is a sharp increase at about $V_{0}\thicksim2t$.
We then identify $V_{0,c}$ as the inflection point of the calculated
curve $\alpha_{\text{IPR}}(V_{0})$ \cite{Dufour2012}, as indicated
by the circle symbol. We have checked that the threshold $V_{0,c}$
is independent on the choice of the phase offset $\theta$. With increasing
disorder strength across $V_{0,c}$, the wave-function of polaron
(impurity) must change from extended to exponentially localized. To
see this, we show in the inset the wave-function of an attractive
polaron in the extended phase ($V_{0}=t$) and of a repulsive polaron
in the localized phase ($V_{0}=3t$). We emphasize that the observed
localization of the polaron wave-function is induced by the impurity-atom
interaction, since the impurity itself does not experience the quasi-disorder
disorder potential. By repeating the calculation of $\alpha_{\text{IPR}}$
for different interaction strengths, we determine the phase boundary
for the localization of the ground-state polaron, as shown in the
phase diagram Fig. \ref{fig1} by solid circles.

\subsection{Many-body localization of all polaron states}

\begin{figure}
\begin{centering}
\includegraphics[clip,width=0.48\textwidth]{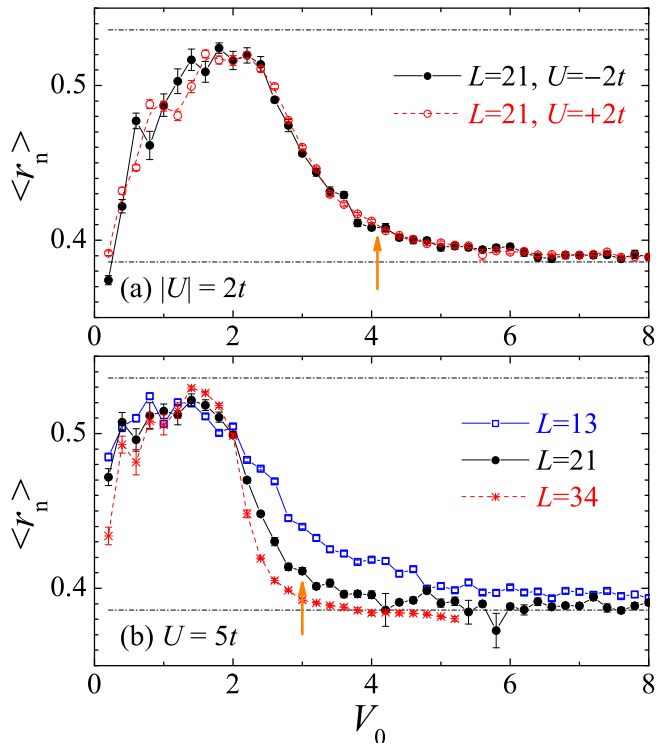} 
\par\end{centering}

\protect\protect\caption{(Color online) Averaged ratio of adjacent energy gaps as a function
of the disorder strength for three values of the interaction strength
$\left|U\right|=2t$ (a) and $U=5t$ (b). The average $\left\langle r_{n}\right\rangle $
was calculated over the central half of the spectrum, averaging over
$20-200$ quasi-random disorder realizations by choosing randomly
the phase offset $\theta$. In (a), we check that the average ratio
is independent on the sign of the interaction strength. In (b), we
show the length dependence of the average ratio. The arrows indicate
the estimated critical disorder strength (see text), at which the
many-body localization of all polaron states occurs. The two thin
dot-dashed lines show the average ratio for the Wigner-Dyson distribution
($\left\langle r_{n}\right\rangle _{\textrm{WD}}\simeq0.536$) and
for the Poisson distribution ($\left\langle r_{n}\right\rangle _{\textrm{P}}\simeq0.386$).}

\label{fig5} 
\end{figure}

We now turn to consider the MBL of all polaron states by computing
the averaged ratio of adjacent energy levels. In Fig. \ref{fig5},
we show the ratio at the interaction strengths $\left|U\right|=2t$
(a) and $U=5t$ (b), as a function of the disorder strength. For any
interaction strength, the disorder dependence of the ratio is similar:
at very weak disorder the ratio approximately takes the value $\left\langle r_{n}\right\rangle _{\textrm{P}}\simeq0.386$
(i.e., the phase I), at some intermediate disorder strengths the ratio
increases to about $\left\langle r_{n}\right\rangle _{\textrm{WD}}\simeq0.536$
(the phase II), and at strong disorder the ratio crosses over to $\left\langle r_{n}\right\rangle _{\textrm{P}}$
again (the phase III). The area of the phase I shrinks quickly by
increasing the absolute value of the interaction strength.

The existence of the phase I can be easily understood from the clean
limit, where the system becomes integrable (or exactly solvable by
Bethe ansatz) \cite{McGuire1965,McGuire1966} and thus loses its ability
to thermalize. The Poisson distribution can be understood as a result
of the localization of the system in momentum space. In contrast,
in the phase III at strong disorder, the system becomes localized
in real space. In between, the system has extended wave-functions
in real space and has the ability to reach thermal equilibrium against
perturbations.

It is readily seen from Fig. \ref{fig5}(a) that the averaged ratio
and hence the MBL do not depend on the sign of the impurity-atom interaction.
The same sign-independence has been experimentally observed for the
localization of a charge-density-wave state \cite{Schreiber2015}.
On the other hand, the averaged ratio depends on the length of the
system, as explicitly shown in Fig. \ref{fig5}(b) for $U=5t$. A
larger system have more Poisson-like statistics than a smaller one
for strong disorder in the apparent localized regime, i.e. $V_{0}>3t$
\cite{Oganesyan2007}. The size dependence of the ratio, however,
becomes weak for a relatively large $L$. In our calculations, we
thus estimate the critical disorder strength of MBL by using the criterion,
\begin{equation}
\left\langle r_{n}\right\rangle _{L=21}\left(V_{0,c}\right)=0.41.
\end{equation}
The uncertainty of this estimation, $\delta V_{0,c}$, can be similarly
determined using the condition, $\left\langle r_{n}\right\rangle _{L=21}\left(V_{0,c}-\delta V_{0,c}\right)=0.43.$
In the figure, the critical disorder strength determined in this manner
has been indicated by the arrow. By repeating the same calculation
for different interaction strengths, we obtain the MBL critical disorder
strength, as reported in the phase diagram Fig. \ref{fig1} by empty
squares.

\section{Conclusions and outlooks}

In summary, we have investigated the many-body localization phenomenon
in the simplest cold-atom setup: a Fermi polaron in quasi-random optical
lattices, where the localization is induced by the impurity-atom interaction.
The use of Chevy's variational approach enables us to access relatively
large samples and therefore we have approximately determined a phase
diagram of many-body localization in the thermodynamic limit. The
localization of the ground-state polaron has also been studied in
greater detail. While at weak disorder both attractive and repulsive
polarons in the ground state behave similarly, at strong disorder
the impurity in an attractive polaron binds with atoms to form a molecule
and the impurity in a repulsive polaron is isolated from atoms. We
note that, both the energy and wave-function of the ground-state polaron
can be experimentally determined by using radio-frequency spectroscopy
and quantum gas microscope, respectively. The ground-state localization
can therefore be directly observed.

Our variational ansatz can be easily generalized to take into account
the effect of the external harmonic trapping potential in real experiments.
Moreover, to improve the quality of the ansatz, we may also consider
\emph{two} particle-hole excitations and use the ansatz\begin{widetext}

\begin{equation}
\left|\text{P2}\right\rangle =\left[\sum\limits _{n}z_{n}d_{n}^{\dagger}+\sum\limits _{n,\eta_{h},\eta_{p}}\alpha_{n}\left(\eta_{h},\eta_{p}\right)\hat{d}_{n}^{\dagger}\hat{c}_{\eta_{p}}^{\dagger}\hat{c}_{\eta_{h}}+\sum\limits _{n,\eta_{h1},\eta_{h2,}\eta_{p1},\eta_{p2}}\alpha_{n}\left(\eta_{h1},\eta_{h2},\eta_{p1},\eta_{p2}\right)\hat{d}_{n}^{\dagger}\hat{c}_{\eta_{p2}}^{\dagger}\hat{c}_{\eta_{p1}}^{\dagger}\hat{c}_{\eta_{h2}}\hat{c}_{\eta_{h1}}\right]\left|\text{FS}\right\rangle .
\end{equation}
\end{widetext}The number of possible states in the enlarged Hilbert
space is about $L^{5}/64$. Therefore, we have $D_{l=6}\simeq5,801$,
$D_{l=7}\simeq63,814$, and $D_{l=8}\simeq709,928$. We may address
the polaron problem with improved accuracy for $l$ up to $8$ and
$L=F_{l}$ up to $34$. 
\begin{acknowledgments}
HH and XJL were supported by the ARC Discovery Projects (Grant Nos.
FT130100815, DP140103231, FT140100003, and DP140100637) and the National
973 program of China (Grant No. 2011CB921502). SY was supported by
the National 973 program of China (Grant No. 2012CB922104) and the
NSFC (Grants Nos. 11434011 and 11421063).\end{acknowledgments}


\begin{thebibliography}{10}
\bibitem{Nandkishore2015}R. Nandkishore and D. A. Huse, Ann. Rev.
Condens. Matter Phys. \textbf{6}, 15 (2015).

\bibitem{Altman2015}E. Altman and R. Vosk, Ann. Rev. of Condens.
Matter Phys. \textbf{6}, 383 (2015).

\bibitem{Basko2006}D. M. Basko, I. L. Aleiner, and B. L. Altshuler,
Ann. Phys. \textbf{321}, 1126 (2006).

\bibitem{Aleiner2010}I. L. Aleiner, B. L. Altshuler, and G. V. Shlyapnikov,
Nature Phys. \textbf{6}, 900 (2010).

\bibitem{Oganesyan2007}V. Oganesyan and D. A. Huse, Phys. Rev. B
\textbf{75}, 155111 (2007).

\bibitem{Pal2010}A. Pal and D. A. Huse, Phys. Rev. B \textbf{82},
174411 (2010).

\bibitem{Bardarson2012}J. H. Bardarson, F. Pollmann, and J. E. Moore,
Phys. Rev. Lett. \textbf{109}, 017202 (2012).

\bibitem{Mondaini2015}R. Mondaini and M. Rigol, Phys. Rev. A \textbf{92},
041601(R) (2015).

\bibitem{Schreiber2015}M. Schreiber, S. S. Hodgman, P. Bordia, H.
P. Lüschen, M. H. Fischer, R. Vosk, E. Altman, U. Schneider, and I.
Bloch, Science \textbf{349}, 842 (2015).

\bibitem{Bordia2015}P. Bordia, H. P. Lüschen, S. S Hodgman, M. Schreiber,
I. Bloch, and U. Schneider, arXiv:1509.00478 (2015).

\bibitem{Harper1955}P. G. Harper, Proc. Phys. Soc. London Sect. A
\textbf{68}, 874 (1955).

\bibitem{Aubry1980}S. Aubry and G. André, Ann. Isr. Phys. Soc. \textbf{3},
133 (1980).

\bibitem{Bloch2008}I. Bloch, J. Dalibard, and W. Zwerger, Rev. Mod.
Phys. \textbf{80}, 885 (2008).

\bibitem{Evers2008}F. Evers and A. D. Mirlin, Rev. Mod. Phys. \textbf{80},
1355 (2008).

\bibitem{Chevy2006} F. Chevy, Phys. Rev. A \textbf{74}, 063628 (2006).

\bibitem{Lobo2006}C. Lobo, A. Recati, S. Giorgini, and S. Stringari,
Phys. Rev. Lett. \textbf{97}, 200403 (2006).

\bibitem{Achirotzek2009}A. Schirotzek, C.-H. Wu, A. Sommer, and M.
W. Zwierlein, Phys. Rev. Lett. \textbf{102}, 230402 (2009).

\bibitem{Wenz2013}A. N. Wenz, G. Zürn, S. Murmann, I. Brouzos, T.
Lompe, and S. Jochim, Science \textbf{342}, 457 (2013).

\bibitem{Massignan2014}P. Massignan, M. Zaccanti, and G. M. Bruun,
Rep. Prog. Phys. \textbf{77}, 034401 (2014).

\bibitem{Cheuk2015}L. W. Cheuk, M. A. Nichols, M. Okan, T. Gersdorf,
V. V. Ramasesh, W. S. Bakr, T. Lompe, and M. W. Zwierlein, Phys. Rev.
Lett. \textbf{114}, 193001 (2015).

\bibitem{Parsons2015}M. F. Parsons, F. Huber, A. Mazurenko, C. S.
Chiu, W. Setiawan, K. Wooley-Brown, S. Blatt, M. Greiner, Phys. Rev.
Lett. \textbf{114}, 213002 (2015).

\bibitem{Omran2015}A. Omran, M. Boll, T. Hilker, K. Kleinlein, G.
Salomon, I. Bloch, and C. Gross, arXiv:1510.04599 (2015).

\bibitem{Doggen2014}E. V. H. Doggen, A. Korolyuk, P. Törmä, and J.
J. Kinnunen, Phys. Rev. A \textbf{89}, 053621 (2014).

\bibitem{Kohmoto1983}M. Kohmoto, Phys. Rev. Lett. \textbf{51}, 1198
(1983).

\bibitem{Dufour2012} G. Dufour and G. Orso, Phys. Rev. Lett. \textbf{109},
155306 (2012).

\bibitem{McGuire1965}J. B. McGuire, J. Math. Phys. \textbf{6}, 432
(1965).

\bibitem{McGuire1966}J. B. McGuire, J. Math. Phys. \textbf{7}, 123
(1966).\end{thebibliography}
\end{document}